# A simple DVH generation technique from various radiotherapy treatment planning systems for independent information system

Byung Jun Min, Heerim Nam, Il Sun Jeong, Hyebin Lee

*Department of Radiation Oncology, Kangbuk Samsung Hospital, Sungkyunkwan University School of Medicine, Seoul 110-746*

In recent years, the use of a picture archiving and communication system (PACS) for radiation therapy has become the norm in hospital environment and has suggested for collecting data and management from different treatment planning systems (TPSs) with the Digital Imaging and Communication in Medicine (DICOM) objects. However, some TPS does not provide the dose-volume histogram (DVH) exportation with text or other format. In addition, plan review systems for various TPSs often allow DVH recalculation with different algorithms. These algorithms result in the inevitable discrepancy between the values obtained with the recalculation and those obtained with TPS itself. The purpose of this study was to develop a simple method for generating reproducible DVH values obtained from the TPSs. Treatment planning information including structures and delivered dose was exported by the DICOM format from Eclipse v8.9 or Pinnacle v9.6 planning systems. The supersampling and trilinear interpolation methods were employed to calculate DVH data from 35 treatment plans. The discrepancies between DVHs extracted from each TPS and the proposed calculation method were evaluated with respect to the supersampling ratio. The volume, minimum dose, maximum dose, and mean dose were compared. The variation of DVHs from multiple TPSs was compared with MIM software v6.1 which is a commercially available treatment planning comparison tool. The overall comparisons of the volume, minimum dose, maximum dose, and mean dose showed that the proposed method generated relatively smaller discrepancies compared with TPS than those by MIM software and TPS. As the structure volume decreased, the overall percent difference increased. Most large difference was observed in the small organs such as eye ball, lens, optic nerve which had below 10 cc volume. A simple and useful technique was developed to generate DVH with acceptable error from a proprietary TPS. This study provides the convenient and common framework which allows to use a single well-managed storage solution for the independent information system.

Email: heerim.nam@samsung.com

Fax: +82-2-2001-1170

## I. INTRODUCTION

Dose-volume histograms (DVHs) are currently used in radiotherapy departments as a significant role in treatment plan evaluation and treatment approval with three-dimensional (3D) dose distributions [1, 2]. Their main benefit to complex treatment plans is providing summarized data of the 3D dose distribution as a graph and statistical format for structures. It is a useful tool for comparison of various treatment plans from different planning techniques or multiple treatment planning systems (TPSs) [3-7]. Moreover, it provides us criteria to overview multi-institutional clinical trials involving advanced delivery technologies [8]. The use of the Digital Imaging and Communications in Medicine (DICOM) and Picture Archiving and Communication System (PACS) has become the norm in the radiotherapy [9-11]. Digitally saved dose data such as dose distribution, DVH, and dose points are widely used for transferring or sharing datasets. However, some treatment systems do not provide DVH exportation with DICOM or other format, and therefore independent DVH calculation algorithm is as necessary as for multiple treatment plan comparisons using different TPS data.

Several studies have shown to generate DVHs based on dose and structure data [2, 12-14]. Typically, DVH is calculated using 3D dose distributions and a shape-based interpolation model for individual structures. The calculation points are chosen by the dose grid and 3D structure delineation through a sampling technique [15]. These interpolation and sampling methods from multiple sources could lead to differences in the DVH calculation. The structure matching via independent methodology could differ from that of the each TPS. Discrete digital image data and different calculation algorithms in individual TPS could also contribute to variations in dose calculation and DVH statistics. Thus, the results of DVH calculation are affected by the structure delineation, sampling resolution, and interpolation algorithm.

In this study, a simple DVH generation method using dose map from various TPSs was developed to reproduce the value obtained with TPS itself. DICOM radiotherapy (RT) structure and dose files were exported from each TPS and utilized for DVH calculation. The statistic data extracted from each TPS and calculated data by proposed method were evaluated with respect to the influence of CT slice thickness and pixel widths, and dose grid widths. The percent discrepancies of minimum, maximum, and mean doses were evaluated by varying the sampling rate of dose grid. All planning data were transferred to a commercial radiation therapy picture archiving and communication system (RTPACS) software as a reference system.

## II. MATERIALS AND METHODS

A. Data collection

Thirty-five patients treated for brain tumors, head and neck cancer, upper abdominal cancer or prostate cancer were selected to evaluate DVH data from different treatment planning systems with 3D conformal radiation therapy (3D-CRT), intensity modulated radiation therapy (IMRT), and volumetric modulated arc therapy (VMAT) techniques. DICOM RT objects including RT structure, RT dose, and RT image were exported from Pinnacle v9.6 (Philipse, Guildford, Surrey, UK) or Eclipse v8.9 (Varian, Palo Alto, CA) treatment planning systems summarized in Table 1. All 3D CT images were acquired using a 16-slice multi-detector CT simulation scanner (LightSpeed RT16, General Electric Healthcare, Milwaukee, WI). CT slice thickness and CT pixel widths were varied to 2.5–5.0 mm and 0.78–1.21 mm, respectively. Among the patients, 19 patients underwent a CT scanning using 2.5 mm slice thickness and the others underwent a CT scanning using 5 mm slice thickness according to our clinical scanning protocol. All CT datasets were transferred to the Pinnacle treatment planning system for the structure contouring.

B. DVH calculation algorithm

DICOM RT structure and DICOM RT dose datasets exported from treatment planning systems were utilized to calculate DVH data. The 3D image and dose volumes are consisted of a set of voxels with a predefined resolution by the user [16]. Since the grid size and its cross-section affect to the DVH calculation, supersampling method, which is a spatial anti-aliasing method, was proposed to improve sampling rate for the calculation [17, 18]. The voxels of the dose dataset were divided into sub-pixels by 2x2, 4x4, 6x6, and 8x8 in transverse direction.

The proposed DVH generation algorithm was calculated according to the Advanced Technology Consortium (ATC) for Clinical Trials Quality Assurance recommendations. The structures were assumed to transverse imaging plane in a set of stacked right prisms. The region of interests (ROI) consisted of a series of closed coplanar axial loops were defined in DICOM RT structures at the center position of CT image. Trilinear interpolation for x, y, and z directions was employed to represent a continuous structure. If the center of each voxel is lying within a structure, it is examined and included in the volume.

The dose statistic values obtained from TPSs were compared to the recalculated values by the proposed supersampling method. The percent differences of minimum dose, maximum dose, and mean dose were evaluated in terms of supersampling ratio. All planning data were transferred to MIM Maestro v6.1 (MIM software Inc., Cleveland, OH) which was used as a reference system for multiple plan comparison. The statistics values were evaluated by comparison with results from each TPS and MIM software.

C. Computation time measurement

Supersampling is computationally expensive because it requires much more computation and memory usage. Therefore, the large structure volumes with fine spatial sampling lead to a heavy computation burden. Calculation time of a relative supersampling rate was measured to evaluate the performance. The measurements of computation time were performed on a general purpose computer with an Intel Core2 Duo CPU (3 GHz) and 4 GB of system memory. The information of structure and dose maps was register in the system memory for the performance enhancement by reducing the memory access time.

## III. RESULTS

A. DVH calculation algorithm

Total 407 structures from 35 treatment plans were analyzed to evaluate a proposed DVH calculation algorithm. Computed structure volumes for various organs were summarized in Table 2. The minimum and maximum volumes for anatomic structures were 0.2 cc (lens) and 1698.7 cc (lung) in median volume. The brain, lung, liver, and bowel showed a large standard deviation. Most structure has the volume of under a hundred cc.

Figure 1 and 2 demonstrate the percent difference between the values of proposed algorithm and those of TPS as a function of the structure volumes. The values of minimum, maximum, and mean doses from TPS were compared with those from proposed calculation method using the different supersampling ratios and with those from MIM software as a reference. As the structure volume decreased, the overall percent difference increased for minimum, maximum, and median doses. These results were similar to those of MIM software. The comparison between the values from TPS and proposed method indicated a good agreement. There were significant differences in the minimum doses. Most large difference was observed in the small organs such as eye ball, lens, optic nerve which had below 10 cc volume. The small absolute values of minimum dose make a big percent discrepancy; some minimum dose from TPS was below 1.0 cGy. Even though percent difference was larger than 3%, the absolute value was below 5. However, the discrepancies in excess of 10% were observed for structure with volume of less than 50 cc.

Table 3 shows the summarized percentage of the overall discrepancy compared to TPS. As the supersampling ratio increased, the discrepancy between recalculated and TPS values was decreased.

DVH based statistics are generally becoming more similar up to a point as sampling resolution increases. The percent discrepancies of dose values obtained custom-made method were improved as compare to those of MIM software.

B. Computation time

The calculation time measurement varying supersampling ratio is demonstrated in Table 4. As supersampling required the more memory and buffer, adaptive supersampling implemented in certain structure area within entire dose grid was used to accelerate calculation speed. The average of three-times measurements for each structure was presented. The calculation time was increased by a factor of supersampling ratio.

## IV. DISCUSSION

In the present study, a simple interpolation method for DVH calculation was developed to improve the reproducibility and data consistency from various TPSs. Since the issue of dose uncertainty from multiple treatment planning systems was not addressed in this study, the main goal was to restore the original DVH data from each TPS. The statistic values obtained by developed DVH calculation using the supersampling method were generally similar to the original values from each TPS. The results demonstrated relatively decreased percent differences by increasing structure volume for minimum, maximum, and median doses. These trends were also similar to the result of comparing the TPS and MIM software. The increased spatial resolution by supersampling was contributed to reduce the uncertainty by interpolation for the matching between structure coordination and dose grid. However, there was no significant improvement for discrepancy even if supersampling ratio was continually increased. Since more sampling required more calculation time, 4x4 sampling showed a reasonable supersampling ratio. Although the different interpolation algorithm resulted in the variation [2], our custom-made DVH generation algorithm demonstrated a good agreement with that of TPS. Most large difference was observed in minimum dose which had below 10 cGy.

DVHs are generally calculated by the combination of a discreet uniform dose grid and structure dataset. Thus, sampling and interpolations are needed to represent a continuous structure matching with the dose and image matrices. Although the structure are sampled and interpolated from the 3D image matrix, it is not exactly matched with the dose grid. Regular or random samplings are used to compute for dose points inside the interested region. This processes result in the uncertainty of dose statistics from various treatment planning systems with different algorithms. In this study, a simple DVH calculation method was proposed to improve the sampling resolution when a dose grid matches with structure sets. Supersampling method was used for more fine resolution near structure edges because most uncertainty was from different interpolation and dose modeling at the boundary of structure. As a result, most discrepancies were observed in the small structures which had many voxels penetrated by contour edges. Because the trilinear interpolation for x, y, z directions was utilized for calculation, there was no dependency according to CT slice thickness matching with the dose grid. Both results using 2.5 mm and 5 mm CT slice thickness were similar.

Typically, the radiation oncology department operates a separate TPS for different modalities. Each system has an independent database in different place to archive treatment plan data with a limited interface for each other. Collecting multiple data sources in a data warehouse combined with information technology solutions is beneficial for clinical research by reducing the time needed to collect necessary data[10]. A proper use of standard protocol is required to effectively communicate between information systems and TPSs from various vendors. The DICOM RT standard, which deals with imaging equipment and PACS, is especially meaningful when specifying data of imaging systems. However, some TPS does not serve the DVH information in DICOM file, and therefore an independent DVH generation method is necessary for analysis of data [2, 13, 14].

Although the drawback of this study was the computation time due to the more sampling, this disadvantage could be overcome by dazzling IT technology. Improved computing power and multi-processing capacity are a sufficient solution for the handicap.

## V. CONCLUSION

In conclusion, a simple DVH calculation method was developed and evaluated for the data consistency to evaluate multiple treatment plans and for the integration to an independent information system. The results demonstrated that the additional spatial sampling of dose grid is useful for restoring the original DVH statistics from various TPSs. This algorithm could be applicable to develop a single well-managed database solution such as RTPACS for an independent information system.

Table 1. Comparison of selected planning data from treatment planning systems (TPSs)

|  | Manufacturer | |
| --- | --- | --- |
|  | Philips | Varian |
| Version | Pinnacle v9.6 | Eclipse v8.9 |
| CT slice thickness (mm) | 2.5–5.0 | 2.5 |
| CT pixel widths (mm) | 0.78–1.21 | 0.78–1.21 |
| Dose grid voxel width (mm) | 2.5–4 | 2.5 |
| DVH dose resolution (cGy) | 5–15 | 1–15 |
| Number of treatment plan | 18 | 7 |

Table 2. Summary of median volume and standard deviation for the computed structure volume

| Structure | Median volume (cc) | Standard deviation (cc) |
| --- | --- | --- |
| GTV | 23.0 | 13.1 |
| CTV | 86.0 | 190.3 |
| PTV | 146.6 | 465.8 |
| Brain Stem | 35.9 | 3.7 |
| Brain | 1319.4 | 201.9 |
| Eye Ball | 9.0 | 1.1 |
| Lens | 0.2 | 0.1 |
| Optic Nerve | 0.8 | 0.3 |
| Parotid | 32.9 | 9.3 |
| Mandible | 71.2 | 22.5 |
| Thyroid | 13.9 | 3.3 |
| Esophagus | 49.2 | 24.3 |
| Spinal Cord | 36.5 | 9.9 |
| Lung | 1698.7 | 586.5 |
| Liver | 1471.4 | 755.1 |
| Trachea | 45.3 | 15.1 |
| Kidney | 192.1 | 37.4 |
| Bowel | 289.7 | 384.0 |
| Duodenum | 33.6 | 6.9 |
| Bladder | 132.5 | 75.1 |
| Rectum | 132.1 | 47.5 |
| Femoral Head | 63.7 | 14.9 |

Table 3. Percentage of percent discrepancy against TPS. Min, max, and mean were minimum, maximum, and mean doses, respectively.

| % discrepancy | Supersampling ratio = 0 | | | Supersampling ratio = 2 | | | Supersampling ratio = 4 | | |
|---|---|---|---|---|---|---|---|---|---|
| | Min | Max | Mean | Min | Max | Mean | Min | Max | Mean |
| ~0.5 | 55.0% | 77.4% | 55.5% | 58.6% | 81.5% | 70.2% | 60.4% | 82.8% | 83.5% |
| 0.5~1 | 13.6% | 8.0% | 18.5% | 13.1% | 6.7% | 18.3% | 10.3% | 8.5% | 6.9% |
| 1~2 | 9.5% | 6.7% | 17.5% | 7.5% | 7.2% | 6.4% | 9.0% | 4.6% | 5.4% |
| 2~4 | 8.2% | 5.4% | 5.1% | 8.0% | 3.6% | 2.8% | 7.2% | 3.1% | 2.8% |
| 4~ | 13.6% | 2.6% | 3.3% | 12.9% | 1.0% | 2.3% | 13.1% | 1.0% | 1.3% |

| % discrepancy | Supersampling ratio = 6 | | | Supersampling ratio = 8 | | | MIM software | | |
|---|---|---|---|---|---|---|---|---|---|
| | Min | Max | Mean | Min | Max | Mean | Min | Max | Mean |
| ~0.5 | 59.6% | 83.0% | 85.1% | 58.6% | 84.3% | 84.1% | 40.4% | 59.4% | 69.9% |
| 0.5~1 | 10.5% | 7.7% | 5.9% | 11.6% | 6.9% | 6.9% | 7.5% | 16.2% | 11.3% |
| 1~2 | 9.5% | 4.4% | 5.4% | 9.5% | 4.4% | 5.4% | 16.2% | 12.3% | 10.3% |
| 2~4 | 7.5% | 3.6% | 2.3% | 6.7% | 3.1% | 2.3% | 9.8% | 6.7% | 6.7% |
| 4~ | 12.9% | 1.3% | 1.3% | 13.6% | 1.3% | 1.3% | 26.2% | 5.4% | 1.8% |

Table 4. Speed measurement results varying supersampling ratio

| | Supersampling ratio | | | | |
|---|---|---|---|---|---|
| | 0 | 2 | 4 | 6 | 8 |
| average computation time (s) | 0.8 | 1.7 | 2.9 | 5.9 | 10.3 |

Figure Captions.

Fig. 1. Comparison of minimum, maximum, and mean dose as a function of structure volume between developed algorithm and TPS with 2.5 mm CT slice thickness.

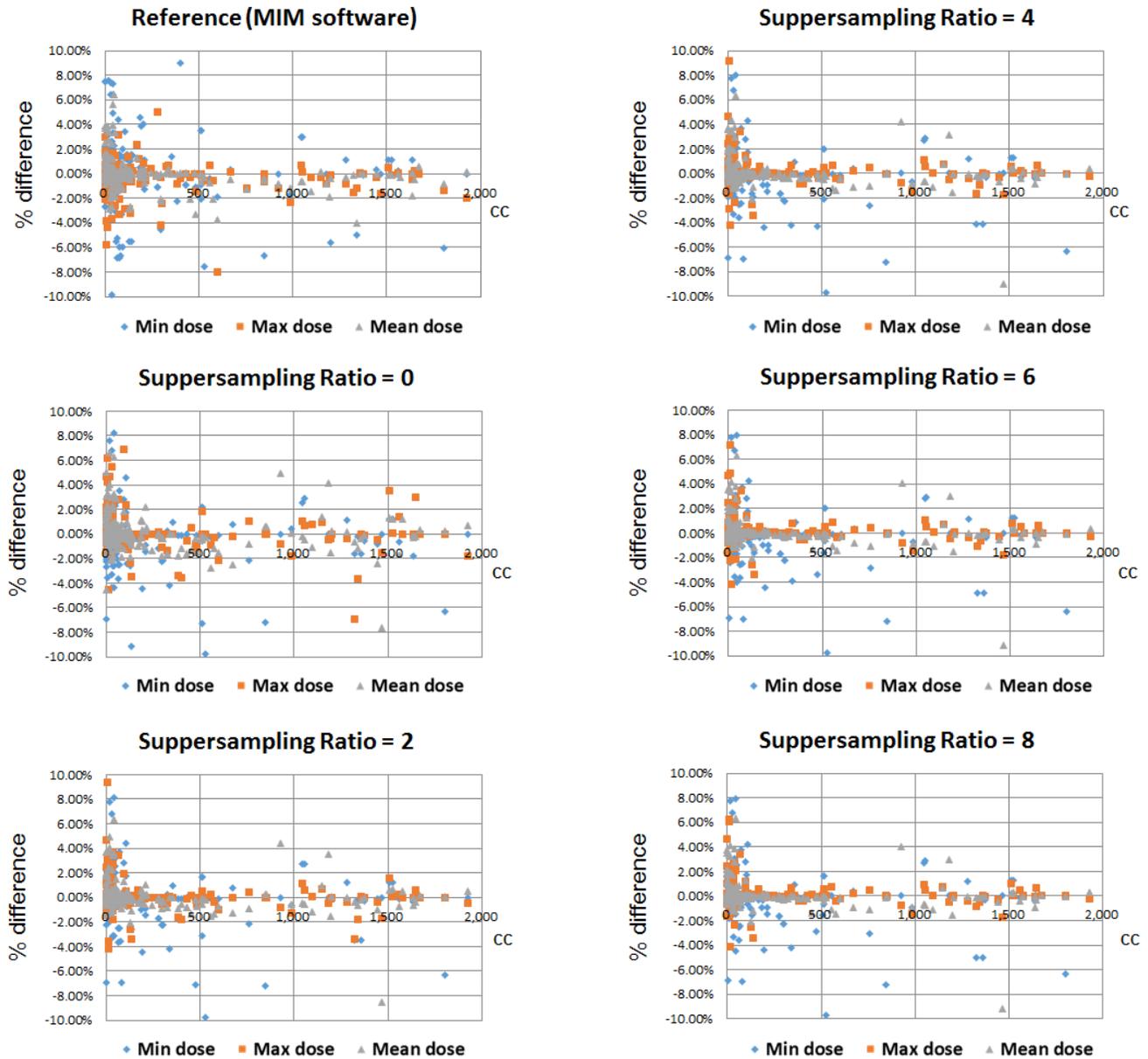

Fig. 2. Comparison of minimum, maximum, and mean dose as a function of structure volume between developed algorithm and TPS with 5 mm CT slice thickness.

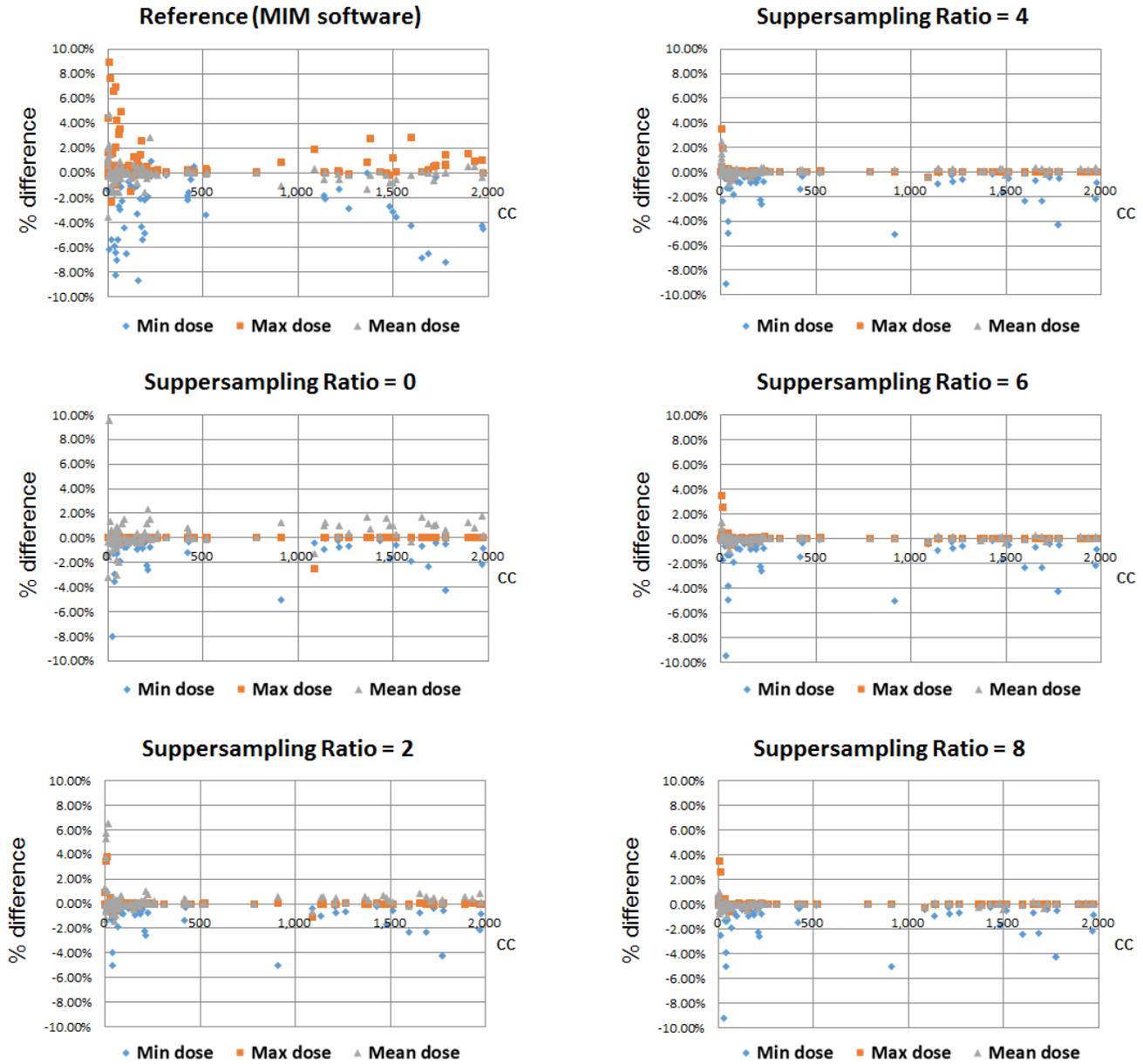